\documentclass[12pt]{article}

\usepackage{graphicx}
\usepackage{amssymb,amsmath}
\usepackage{subfigure}
\usepackage{color}
\usepackage{mdwlist}

\graphicspath{{./figures/}}

\newcommand{\captionfonts}{\sf}

\makeatletter  
\long\def\@makecaption#1#2{%
  \vskip\abovecaptionskip
  \sbox\@tempboxa{{\captionfonts #1: #2}}%
  \ifdim \wd\@tempboxa >\hsize
    {\captionfonts #1: #2\par}
  \else
    \hbox to\hsize{\hfil\box\@tempboxa\hfil}%
  \fi
  \vskip\belowcaptionskip}
\makeatother   

\makeatletter

\renewcommand{\familydefault}{}

\def\ve#1{{\mathchoice{\mbox{\boldmath$\displaystyle #1$}}%
              {\mbox{\boldmath$\textstyle #1$}}%
              {\mbox{\boldmath$\scriptstyle #1$}}%
              {\mbox{\boldmath$\scriptscriptstyle #1$}}}}

\newcommand{\evs}{\mathcal{E}}

\newcommand{\Ex}{\mathbb{E}}

\newcommand{\SNR}{\bar{E}_b / \mathcal{N}_0}

\newcommand{\diag}{\mathrm{diag}}
\newcommand{\PEP}{{\mathrm{PEP}}}
\newcommand{\No}{\mathcal{N}_0}

\newcommand{\SIR}{\mathrm{SIR}}

\def\section{\@startsection{section}{1}{\z@}
     {-1.5ex plus-1ex minus -.2ex}{2.0ex minus 2.0ex}
     {\Large\sf\textbf}}

\def\subsection{\@startsection{subsection}{2}{\z@}
     {-2.5ex plus-1ex minus -.2ex}{1.5ex plus.2ex}
     {\large\sf\textbf}}
\def\subsubsection{\@startsection{subsubsection}{3}{\z@}
     {-1.25ex plus-1ex minus -.2ex}{1.5ex plus.2ex}
     {\normalsize\sf\textbf}}
\def\paragraph{\@startsection{paragraph}{4}{\z@}
     {-1.25ex plus-1ex minus -.2ex}{1.5ex plus.2ex}
     {\normalsize\sf\textsl}}

\newcommand{\Section}[1]{\vspace{-5mm}\section{#1}\vspace{-2mm}} 
\newcommand{\SubSection}[1]{\vspace{-5mm}\subsection{#1}\vspace{-3mm}} 
\newcommand{\SubSubSection}[1]{\vspace{-2mm}\subsubsection{#1}\vspace{-4mm}}

\def\ps@headings{\let\@mkboth=\markboth
        \def\@oddhead{\vbox{\hsize\textwidth \hbox to \textwidth{%
                        \small\sf 
                        {Snow \textsl{et al.}: Error Rate Analysis for Coded Multicarrier Systems over Quasi-Static Fading Channels}%
                        \hfil \thepage}
                        \vskip 3pt \hrule}\hss%
        }
        \def\@oddfoot{\hss%
        }
}
\makeatother

\begin{document}
\sf

\thispagestyle{empty}
\vspace*{-30mm}
\begin{center}
{\Huge Error Rate Analysis for Coded Multicarrier Systems over Quasi-Static Fading Channels$^*$}
\end{center}
\vspace*{5mm}

\begin{center}
Chris Snow$^\dagger$, Lutz Lampe, and Robert Schober \\[0.5em]
Department of Electrical and Computer Engineering \\
The University of British Columbia \\
Vancouver, British Columbia, Canada \\
Email: $\{$csnow, lampe, rschober$\}$@ece.ubc.ca
\end{center}
\vspace*{6mm}

\renewcommand{\baselinestretch}{1.5}
\large\normalsize

{{\slshape Abstract}} --- Several recent standards such as IEEE 802.11a/g, IEEE 802.16, and ECMA Multiband Orthogonal Frequency Division Multiplexing (MB-OFDM) for high data-rate Ultra-Wideband (UWB), employ bit-interleaved convolutionally-coded multicarrier modulation over quasi-static fading channels. Motivated by the lack of appropriate error rate analysis techniques for this popular type of system and channel model, we present two novel analytical methods for bit error rate (BER) estimation of coded multicarrier systems operating over frequency-selective quasi-static channels with non-ideal interleaving. In the first method, the approximate performance of the system is calculated for each realization of the channel, which is suitable for obtaining the outage BER performance (a common performance measure for e.g.~MB-OFDM systems). The second method assumes Rayleigh distributed frequency-domain subcarrier channel gains and knowledge of their correlation matrix, and can be used to directly obtain the average BER performance. Both methods are applicable to convolutionally-coded interleaved multicarrier systems employing Quadrature Amplitude Modulation (QAM), and are also able to account for narrowband interference (modeled as a sum of tone interferers). To illustrate the application of the proposed analysis, both methods are used to study the performance of a tone-interference-impaired MB-OFDM system.

\vspace*{5mm}


\vspace*{5mm}
\renewcommand{\baselinestretch}{1.0}
\large\normalsize

$^*$ {\small This work has been submitted in part for presentation at the 2006 IEEE Global Telecommunications Conference (GLOBECOM), and has been accepted in part for presentation at the 2006 IEEE International Conference on Ultra-Wideband (ICUWB). This work has been supported in part by the National Science and Engineering Research Council of Canada (Grant CRDPJ 320552) and Bell University Laboratories, and in part by a Canada Graduate Scholarship.}

$^\dagger$ {\small Corresponding author}

\newpage
\renewcommand{\baselinestretch}{1.66}\normalsize
\setcounter{page}{1}

\Section{Introduction}

Multicarrier communication systems based on Orthogonal Frequency Division Multiplexing (OFDM) have gained interest from the communications community in recent years, as evidenced by standards such as xDSL (digital subscriber lines), IEEE 802.11a/g (wireless local area networks) \cite{802.11}, IEEE 802.16 (broadband wireless access) \cite{KR02}, and ECMA Multiband OFDM (MB-OFDM) for high-rate Ultra-Wideband (UWB)~\cite{ECMA,MBOFDM}. In general, the channel for these systems can be assumed to be very slowly time-varying relative to the transmission rate of the device, and can be approximated as quasi-static for the duration of one or more packet transmissions. As well, since most OFDM devices use a relatively large bandwidth, the channel is frequency selective. Simulation-based approaches to obtain system performance in this setting are very time consuming due to the necessity of simulating the system over a large number of channel realizations. 

The complexity of simulation and the wide range of practical systems operating over this type of channel motivate the relevance of, and interest in, analytical methods for obtaining the performance of coded multicarrier systems when transmitting over quasi-static frequency-selective fading channels. There are well-known techniques for bounding the performance of convolutionally-encoded transmission over many types of fading channels, e.g.~\cite{Proakis,CaireEtAl98}. However, such classical bit error rate (BER) analysis techniques are not applicable to the OFDM systems mentioned above for several reasons. Firstly, the short-length channel-coded packet-based transmissions are non-ideally interleaved, which results in non-zero correlation between adjacent coded bits. Secondly, and more importantly, the quasi-static nature limits the number of distinct channel gains to the (relatively small) number of OFDM tones. This small number of distinct channel gains must not be approximated by the full fading distribution for a valid performance analysis, as would be the case in a fast-fading channel.

In the quasi-static channel setting it is also often of significant interest to obtain the outage BER performance, i.e., the minimum expected BER performance after excluding some percentage of the worst-performing channel realizations~\cite[Section~III.C-2]{BPS98}. Until now, one has had to resort to intensive numerical simulations in order to obtain the BER performance for each channel realization, and hence obtain the outage BER performance \cite{BatraEtAl04}. 

Motivated by the considerations mentioned above, we have developed two analytical methods to obtain outage as well as average BER performance over ensembles of channel realizations. The two methods are best-suited to different types of analysis, as will be discussed in detail below.

Furthermore, in the UWB setting, OFDM-based schemes are used as spectral underlay systems for previously allocated frequency bands. The resultant interference in this setting motivates the investigation of the effect of narrowband interference on OFDM systems. While narrowband interference models have been considered for impulse radio and direct-sequence UWB systems \cite{GCW05}, there are no results available in the open technical literature on the effect of narrowband interference on coded MB-OFDM.

\textsl{Contributions:} We briefly outline the main contributions of this paper:
\vspace{-12pt}
\begin{itemize*}
\item We develop a method for approximating the BER of coded multicarrier systems on a \textsl{per-realization} basis (``Method I''). This method is most suitable for obtaining the outage BER, but can also be used to obtain the average BER performance (Section~\ref{sec:rba}).
\item For quasi-static channels with correlated Rayleigh-distributed subcarrier channel gains, we present an alternative method (``Method II'') to directly and efficiently obtain the average BER performance (Section~\ref{sec:avg}).
\item We model narrowband interference as a sum of tone interferers (a reasonable model for evaluating the effect of one or more interferers with narrow bandwidth as compared to the OFDM subcarrier spacing), and incorporate the effects of this interference into both analysis methods.
\item As a specific example, and motivated by the inherent interference in UWB communications, we provide numerical results illustrating the two BER approximation methods applied to the MB-OFDM UWB system. In particular, we present both outage and average BER, with and without tone interference (Section~\ref{sec:results}).
\item Erasure marking and decoding \cite{DP02,LMS03} is a promising technique to alleviate the effects of interference. We study erasure marking and decoding as a mitigation technique for tone-interference-impaired coded MB-OFDM (Section~\ref{sec:results_erasures}).
\end{itemize*}

\textsl{Related Work:} In \cite{ML99}, Malkam\"aki and Leib consider the performance of convolutional codes with non-ideal interleaving over \textsl{block} fading channels without interference. They make use of the generalized transfer function (GTF) \cite{LWK93} method in order to obtain the pairwise error probability (PEP). If their technique is applied to systems with a fading block length of one (equivalent to the quasi-static channel), their approach is similar in some ways to Method~I presented in Section~\ref{sec:rba}. The major difference is that Method~I does not require the GTF of the code, which may become difficult to obtain as the number of distinct channel gains (the number of blocks in the case of a block-fading channel) grows~\cite{CCT04}. 
Instead, we apply the novel concept of error vectors, introduced in Section~\ref{sec:evs}.

The PEP for uncoded and coded (across subcarrier) MB-OFDM is given in \cite{SSL06}. However, the authors apply a non-standard UWB channel model, consider only simple codes such as repetition coding, and do not consider interference.

\textsl{Organization:} The remainder of this paper is organized as follows. Section~\ref{sec:models} introduces the OFDM transmitter and receiver models as well as the models for the channel and for the interfering signals. Each model is formulated quite generally, although we also mention the specific parameters for the MB-OFDM system, which will be the focus of numerical results presented. In Section~\ref{sec:analysis}, we develop the proposed analysis methods, which allow for per-channel-realization as well as average error rate approximations with and without sum-of-tones interference. Analysis and simulation results for several practically relevant scenarios of interest for MB-OFDM are given and discussed in Section~\ref{sec:results}. Finally, Section~\ref{sec:conclusions} concludes this paper.

\textsl{Notation:} In this paper, $(\cdot)^T$ and $(\cdot)^H$ denote transpose and Hermitian transpose, respectively, $\diag(\ve{x})$ denotes a matrix with the elements of vector $\ve{x}$ on the main diagonal, $\mathrm{Re}\{\cdot\}$ denotes the real part of a complex number, $\Ex(\cdot)$ denotes the expectation of a random variable, $\mathrm{Pr}\{\cdot\}$ denotes the probability of some event, $\mathrm{DFT}(\cdot)$ denotes the Discrete Fourier Transform, $Q(\cdot)$ is the Gaussian Q-function~\cite{Proakis}, and $\oplus$ denotes the element-wise XOR operation. Matrices $\ve{I}_\eta$ and $\ve{0}_\eta$ denote the identity matrix and the all-zero matrix of dimension $\eta\times\eta$, respectively, $\ve{0}_{\eta{\times}1}$ denotes the all-zero column vector of length $\eta$, and $\det(\cdot)$ denotes the determinant of a matrix.

\Section{System Model}
\label{sec:models}

In this section, we introduce the OFDM transmitter, the channel and interference models, and the OFDM receiver. We again note that the methods presented herein are applicable to the general class of coded multicarrier systems operating over quasi-static fading channels, although we have chosen to focus our attention on the MB-OFDM system for high data-rate UWB.

\SubSection{Transmitter}

\begin{figure}[t]
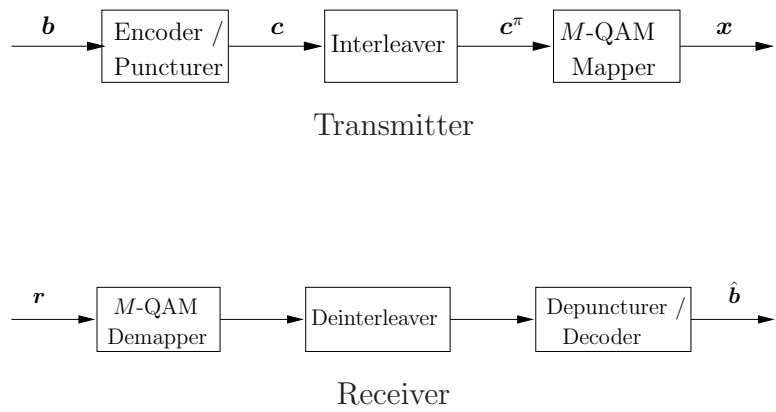

\centering
        \subfigure{
        \label{fig:sysdiag_transmitter}
        \resizebox{.6\columnwidth}{!}{\input{figures/transmitter.pstex_t}} }
        \centerline{Transmitter}
        \vspace{1cm}

        \subfigure{
        \label{fig:sysdiag_receiver}
        \resizebox{.6\columnwidth}{!}{\input{figures/receiver.pstex_t}} }
        \centerline{Receiver}
        \vspace{1cm}
        \caption{\label{fig:sysdiag} %
        Relevant portions of the OFDM transmission system.}
\end{figure}

Throughout this paper we consider an $N$-subcarrier OFDM system with $M$-ary QAM ($M$-QAM) carrying $R_m=\log_2(M)$ bits per subcarrier. Figure~\ref{fig:sysdiag} shows the relevant portions of the OFDM transmitter. The system employs a punctured convolutional code of rate $R_c$. 
We assume that the transmitter selects $R_cR_mN$ random message bits for transmission, denoted by $\ve{b} = [b_1\;b_2 \dots b_{R_cR_mN}]^T$. The vectors $\ve{c}$ and $\ve{c}^\pi$ of length $L_c=R_mN$ represent the bits after encoding/puncturing and after interleaving, respectively. The bits $\ve{c}^\pi$ are then modulated using $M$-QAM on each subcarrier, and the resulting $N$ modulated symbols are denoted by $\ve{x} = [x_1\;x_2\;\dots\;x_N]^T$.

As a specific example, we consider the MB-OFDM system proposed for the IEEE 802.15 TG3a high data-rate UWB standard and recently standardized by the ECMA~\cite{ECMA,MBOFDM,BatraEtAl04}. The MB-OFDM system employs 128 subcarriers, and hops over 3 sub-bands (one hop per OFDM symbol) for first-generation devices. We assume without loss of generality that hopping pattern 1 of \cite{MBOFDM} is used (i.e.~the sub-bands are hopped in order). As a result we can consider MB-OFDM as an equivalent 384 subcarrier OFDM system. After disregarding pilot, guard, and other reserved subcarriers, we have $N=300$ data-carrying subcarriers.

Channel coding in the proposed standard consists of a punctured maximum free distance rate $1/3$ constraint length 7 convolutional encoder and a multi-stage block-based interleaver (see \cite{MBOFDM} for details). After modulation, modulated symbols are optionally repeated in two consecutive OFDM symbols and/or two subcarriers within the same OFDM symbol, reducing the effective code rate by a factor of 2 or 4 and providing an additional spreading gain for low data rate modes. We can equivalently consider this repetition as a lower-rate convolutional code with repeated generator polynomials. The proposed standard specifies only 4-QAM with Gray labeling. We also consider Gray-labeled 16-QAM as a potential extension for increased data rates, and since it is also used in other OFDM-based standards such as IEEE 802.11a/g and IEEE 802.16.

\SubSection{Channel Model}
\label{sec:channelmodel}

We will assume that the OFDM system is designed such that the cyclic prefix is longer than the channel impulse response. Thus, we can equivalently consider the channel in the frequency domain, and denote the subcarrier gains by $\ve{h} = [h_1\;h_2\;\dots\;h_N]$. We also include the frequency-domain interference signal $\ve{J}$  (see Section~\ref{sec:interferencemodel}). The transmitted symbols $\ve{x}$ pass through the fading channel $\ve{H}=\diag(\ve{h})$, and the length-$N$ vector of received symbols $\ve{r}$ (after the DFT) is given by
\begin{equation}
\label{eq:rxmodel}
\ve{r} = \ve{H}\ve{x} + \ve{J} + \ve{n}\;,
\end{equation}
where $\ve{n}$ is a vector of independent complex additive white Gaussian noise (AWGN) variables with variance $\mathcal{N}_0$. We denote the energy per modulated symbol by $E_s=R_cR_mE_b$, where $E_b$ is the energy per information bit.

For a meaningful performance analysis of the MB-OFDM proposal, we consider the channel impulse response model developed under IEEE 802.15 for UWB systems \cite{MolischEtAl03} (a version of the Saleh-Valenzuela model~\cite{Saleh+Valenzuela87} modified to fit the properties of measured UWB channels). Multipath rays arrive in clusters, with exponentially distributed cluster and ray interarrival times. Both clusters and rays have decay factors chosen to meet a given power decay profile. The ray amplitudes are modeled as lognormal random variables, and each cluster of rays also undergoes a lognormal fading. To provide a fair system comparison, the total multipath energy is normalized to unity. As well, the entire impulse response undergoes an ``outer'' lognormal shadowing.  The channel impulse response is assumed time invariant during the transmission period of (at least) one packet. We consider the UWB channel parameter set CM1 for short-range line-of-sight channels~\cite{MolischEtAl03}. We also note that, due to the relative complexity of this channel model, even BER analysis for uncoded UWB systems has proven challenging~\cite{GH05}.

The outer lognormal shadowing term mentioned above is irrelevant for the fading characteristics as it  affects all tones equally. Denoting the lognormal term by $G$, we obtain the corresponding \textsl{normalized} frequency-domain fading coefficients as 
\begin{equation}
\label{eq:normchannel}
        \ve{h}^n = \ve{h}/G\;.
\end{equation}

The elements of $\ve{h}^n$ are well-approximated as zero-mean complex Gaussian random variables~\cite{SLS05a,Wessman04}. This allows us to apply analysis assuming correlated Rayleigh fading coefficients (see Section~\ref{sec:avg}) to the UWB channel without lognormal shadowing, and then average over the lognormal shadowing distribution in order to obtain the final system performance over the UWB channel. We note that this is only relevant for the method in Section \ref{sec:avg} --- for the realization-based method (see Section~\ref{sec:rba}) the distribution of $\ve{h}$ is not important.

\SubSection{Interference Model}
\label{sec:interferencemodel}

We model narrowband interference as the sum of $N_i$ tone interferers
\begin{equation}
i(t) = \sum_{k=1}^{N_i}i_{k}(t) \; ,
\end{equation}
where the equivalent complex baseband representation of the $k$th tone interferer with amplitude $\alpha_k$, frequency $f_k$, and initial phase $\phi_k$ is given by
\begin{equation}
i_k(t) = \alpha_ke^{j(2{\pi}{f_k}t+\phi_k)} \; .
\end{equation}

Assuming that the interference $i(t)$ falls completely in the passband of the receiver filter before sampling, we form the discrete-time equivalent interference by sampling $i(t)$ with the OFDM system sampling period $T$, and obtain (for one OFDM symbol) the $N$ sample vector
\begin{equation}
\ve{i} = [ i(0) \quad i(T) \quad i(2T) \quad \dots \quad i((N-1)T) ]^T \; .
\end{equation}

Therefore, the frequency-domain equivalent $\ve{J}$ of the interfering signal considered in (\ref{eq:rxmodel}) is given by
\begin{equation}
\ve{J} = \mathrm{DFT}( \ve{i} ) \; .
\end{equation}

We note that, due to the finite-length DFT window, each single-tone interferer is convolved by a sinc-function in the frequency domain. If $f_k$ is equal to one of the subcarrier frequencies, only one subcarrier is impaired by the interferer $i_k(t)$ (since the interferer will be zero at the other subcarrier frequencies). On the other hand, if $f_k$ happens to lie between two subcarriers, the tone interferer will affect several adjacent subcarriers.

\SubSection{Receiver}

The relevant portions of the OFDM receiver are shown in Figure~\ref{fig:sysdiag}. We assume perfect timing and frequency synchronization. The receiver employs a soft-output detector followed by a deinterleaver and a depuncturer. After possible erasure marking based on knowledge of $f_k,1\;{\le}\;k\;{\le}\;N_i$ (see Section~\ref{sec:results_erasures} for details), standard Viterbi decoding results in an estimate $\hat{\ve{b}} = [\hat{b}_1\;\hat{b}_2\;\dots\;\hat{b}_{R_cR_mN}]^T$ of the original transmitted information bits.

\Section{Performance Analysis}
\label{sec:analysis}

In this section, we present two methods for approximating the performance of coded multicarrier systems operating over frequency-selective, quasi-static fading channels and impaired by sum-of-tones interference. The first method (Section \ref{sec:rba}) is based on approximating the performance of the system for individual channel realizations. The main strength of this method is that it can be used to obtain the outage BER performance (the standard performance measure considered in MB-OFDM systems \cite{ECMA,MBOFDM,BatraEtAl04}). While the first method can also be used to obtain the average BER over an ensemble of channel realizations, the second method (Section \ref{sec:avg}), which is based on knowledge of the correlation matrix of the frequency-domain channel gains, can be used to directly obtain the average performance without the need to consider a large ensemble of channels. Both methods are based on considering the set of error vectors, introduced in Section~\ref{sec:evs}, and the PEP of an error vector, given in Section~\ref{sec:ev_pep}.

One major problem in the analysis of $M$-QAM schemes with $M>4$ is that the probability of error for a given bit depends on the whole transmitted symbol (i.e., it also depends on the other bits in the symbol). For this reason, for the combination of convolutional coding and $M$-QAM it is not sufficient to adopt the classical approach of considering deviations from the all-zero codeword only. In theory, one must average over all possible choices for $\ve{c}$. Since this is computationally intractable, we simply assume the transmitted information bits $\ve{b}$ (and hence $\ve{x}$) are chosen randomly. For $M=4$ (where the joint linearity of code and modulator is maintained) this is exactly equivalent to considering an all-zero codeword. In the case of $M>4$, we have verified for various example scenarios that, for the two analysis methods proposed below, a random choice of $\ve{b}$ well-approximates the true system performance.

\SubSection{Error Vectors}
\label{sec:evs}

Let $\evs$ be the set of all $L$ vectors  $\ve{e}_\ell$ $(1\;{\le}\;\ell\;{\le}\;L)$ of code output (after puncturing) associated with input sequences whose Hamming weight is less than $w_{\mathrm{max}}$, i.e., $\evs = \left\{\ve{e}_1,\ve{e}_2,\dots,\ve{e}_L\right\}$. We limit $\evs$ to only \textsl{simple} vectors \cite{SD03}, where the error path contains only one deviation from (and return to) the all-zero trellis state. Let $l_\ell$ and $a_\ell$ be the length of $\ve{e}_\ell$ and the number of information bit errors associated with $\ve{e}_\ell$, respectively. As with standard union-bound techniques for convolutional codes~\cite{Proakis}, the low-weight terms will dominate the error probability. Hence, it is sufficient to choose a small $w_{\mathrm{max}}$ --- for example, the punctured MB-OFDM code of rate $R_c=1/2$ \cite{MBOFDM} has a free distance of $9$, and choosing $w_{\mathrm{max}}=14$ (resulting in a set of $L=242$ error vectors of maximum length $l=60$) provides results virtually identical to those obtained using larger $w_{\mathrm{max}}$ values.

We term $\ve{e}_\ell$ an ``error vector'' and $\evs$ the set of error vectors. Note that $\evs$ can be straightforwardly obtained from the transfer function of the code, without resorting to the GTF~\cite{LWK93} approach adopted in~\cite{ML99}, and is also independent of the number of distinct channel gains (or the number of blocks in the context of~\cite{ML99}).

\SubSection{Pairwise Error Probability (PEP) for an Error Vector}
\label{sec:ev_pep}

We consider error events starting in a given position $i$ of the codeword ($1 \le i \le L_c$). For a specific error vector $\ve{e}_\ell$ ($1 \le \ell \le L$), form the full error codeword 
\begin{equation}
\label{eq:q}
\ve{q}_{i,\ell} = [\underbrace{0\phantom{_\ell}\;0\;\dots\;0}_{i-1}\;\underbrace{\ve{e}_\ell}_{l_\ell}\;\underbrace{0\phantom{_\ell}\;0\;\dots\;0}_{L_c-l_\ell-i+1}]^T
\end{equation}
of length $L_c$ by padding $\ve{e}_\ell$ with zeros on both sides as indicated above. Given the error codeword $\ve{q}_{i,\ell}$ and given that codeword $\ve{c}$ is transmitted, the competing codeword is given by
\begin{equation}
\label{eq:v}
\ve{v}_{i,\ell} = \ve{c} \oplus \ve{q}_{i,\ell} \quad.
\end{equation}
The decoder employs a standard Euclidean distance metric (i.e., the interference signal is assumed to be unknown for calculation of the metric). Letting $\ve{z}_{i,\ell}$ be the vector of $M$-QAM symbols associated with $\ve{v}_{i,\ell}^\pi$ (the interleaved version of $\ve{v}_{i,\ell}$), and recalling that $\ve{x}$ is the modulated symbol vector corresponding to the original codeword $\ve{c}$, the PEP for the $\ell$th error vector starting in the $i$th position, i.e., the probability that $\ve{v}_{i,\ell}$ is detected given that $\ve{c}$ was transmitted, is given by
\begin{equation}
\label{eq:PEP}
\PEP_{i,\ell}(\ve{H},\ve{J}) = \mathrm{Pr}\left\{||\ve{r} - \ve{H}\ve{x}||^2 > ||\ve{r}-\ve{H}\ve{z}_{i,\ell}||^2 \; \big| \; \ve{H},\ve{J} \right\} \;.
\end{equation}

In Sections \ref{sec:rba} and \ref{sec:avg}, we will obtain various forms for this general expression.

\SubSection{Per-realization Performance Analysis (``Method I'')}
\label{sec:rba}

In this section, we obtain an approximation of the BER for a particular channel realization $\ve{H}=\diag(\ve{h})$ and interference $\ve{J}$, which we denote as $P(\ve{H},\ve{J})$. For simplicity, we refer to this method as ``Method I'' in the remainder of this paper. As noted above and discussed in more detail in Section \ref{sec:AvgOutBER}, the main strength of this method is the ability to obtain the outage BER of coded OFDM systems.

\SubSubSection{Pairwise Error Probability (PEP)}

The PEP for an error vector $\ve{e}_\ell$ ($1 \le \ell \le L$) with the error event starting in a position $i$ ($1 \le i \le L_c$) is given by (\ref{eq:PEP}).  For a given $\ve{H}$ and $\ve{J}$, and after some straightforward manipulations, we obtain the expression
\begin{equation}
\label{eq:PEPij}
\PEP_{i,\ell}(\ve{H},\ve{J}) = Q \left( \frac{\frac{1}{2}||\ve{H}(\ve{x}-\ve{z}_{i,\ell})||^2 + \mathrm{Re}\left\{\ve{J}^{H}\ve{H}(\ve{x}-\ve{z}_{i,\ell})\right\}} %
    {\sqrt{\frac{1}{2}\No||\ve{H}(\ve{x}-\ve{z}_{i,\ell})||^2}} \right) \; .
\end{equation}

It is insightful to examine (\ref{eq:PEPij}) for two special cases:
\begin{itemize}
\vspace{-12pt}
\item $\No\rightarrow0$ (the low-noise region): In this case, there are two possible outcomes. If the numerator in (\ref{eq:PEPij}) is positive, we have the $Q$-function of a large positive value and thus $\PEP\rightarrow0$. However, if the interference $\ve{J}$ causes the numerator to become negative, we have the $Q$-function of a large negative value and thus $\PEP\rightarrow1$. That is, we either (depending on $\ve{J}$) will surely make an error, or will surely \textsl{not} make an error.
\vspace{-12pt}
\item $\ve{J}=\ve{0}_{N{\times}1}$ (no interference): Here we can simplify (\ref{eq:PEPij}) to obtain
\begin{equation}
\label{eq:PEPnoI}
\PEP_{i,\ell}(\ve{H},\ve{J}) = Q\left(\sqrt{\frac{||\ve{H}(\ve{x}-\ve{z}_{i,\ell})||^2}{2\No}}\right) \;.
\end{equation}
\end{itemize}

\SubSubSection{Per-realization BER}

The corresponding bit error rate for the $\ell$th error vector, starting in the $i$th position, is given by
\begin{equation}
\label{eq:BERij}
P_{i,\ell}(\ve{H},\ve{J}) = a_\ell \cdot \PEP_{i,\ell}(\ve{H},\ve{J}) \;.
\end{equation}
Summing over all $L$ error vectors, we obtain an approximation of the BER for the $i$th starting position as
\begin{equation}
\label{eq:BERi}
P_i(\ve{H},\ve{J}) = \sum_{\ell=1}^{L}P_{i,\ell}(\ve{H},\ve{J}) \;.
\end{equation}
We note that (\ref{eq:BERi}) can be seen as a standard truncated union bound for convolutional codes (i.e., it is a sum over all error events of Hamming weight less than $\omega_{\mathrm{max}}$). We also note that we can tighten this bound by limiting $P_i$ to a maximum value of $1/2$ before averaging over starting positions~\cite{ML99}. Finally, since all starting positions are equally likely, the BER $P(\ve{H},\ve{J})$ can be written as 
\begin{equation}
\label{eq:BERgivenH}
P(\ve{H},\ve{J}) = \frac{1}{L_c}\sum_{i=1}^{L_c}\min \left[ \frac{1}{2}, \sum_{\ell=1}^{L} P_{i,\ell}(\ve{H},\ve{J}) \right]\;.
\end{equation}
Table~\ref{table:pseudo} contains pseudocode to calculate $P(\ve{H},\ve{J})$ according to (\ref{eq:BERgivenH}).

\SubSubSection{Average and Outage BER}
\label{sec:AvgOutBER}

The average BER for a given interference can be obtained by averaging (\ref{eq:BERgivenH}) over a (large) number $N_c$ of channel realizations, where the $i$th channel realization is denoted by $\ve{H}_{i}$ ($1 \le i \le N_c$), as
\begin{equation}
\label{eq:BERavgRealiz}
P(\ve{J}) = \frac{1}{N_c}\sum_{i=1}^{N_c}P(\ve{H}_{i},\ve{J})\;.
\end{equation}

As mentioned previously, Method I also readily lends itself to the consideration of the outage BER, a common measure of performance for packet-based systems operating in quasi-static channels \cite{BPS98}. The outage BER provides a measure of the minimum performance that can be expected of the system given a specified $X$\% outage rate, and is often employed in UWB system performance studies \cite{BatraEtAl04}. We evaluate (\ref{eq:BERgivenH}) for a set of $N_c$ channel realizations $\mathcal{H}=\{\ve{H}_i, 1 \le i \le N_c \}$. The worst-performing $X$\% of realizations are considered in outage, and those channel realizations are denoted by $\mathcal{H}_{\mathrm{out}}$. Denoting the remaining $(100-X)$\% of channel realizations by $\mathcal{H}_{\mathrm{in}}$, the outage BER is given by
\begin{equation}
\label{eq:BERoutRealiz}
P_{\mathrm{out}}(\ve{J}) = \max_{\ve{H}_i \in \mathcal{H}_{\mathrm{in}}} P(\ve{H}_i,\ve{J})\;.
\end{equation}
In Section~\ref{sec:results}, we will focus on results for fixed values of signal-to-interference ratio ($\SIR$), interferer amplitude $\alpha_k$, and interferer frequency $f_k$. However, in order to remove the effect of the interferer initial phase, we will average (\ref{eq:BERavgRealiz}) and (\ref{eq:BERoutRealiz}) over 32 uniformly-distributed values of $\phi_k \in [0,2\pi)$.

\SubSection{Average Performance Analysis (``Method II'')}
\label{sec:avg}

In this section, we propose a method, based on knowledge of the frequency-domain channel correlation matrix, which can be used directly in order to obtain the average BER performance of coded multicarrier systems. The advantage of this method is that it allows for simple and direct evaluation of the average BER, without the need to evaluate the BER of many different channel realizations as in Method~I, cf. (\ref{eq:BERavgRealiz}). For simplicity, we refer to this method as ``Method II'' in the remainder of this paper.

For this method we will explicitly assume that the elements of $\ve{h}$ are Rayleigh-distributed and have known correlation matrix $\ve{\Sigma_{hh}}$ (in practice, $\ve{\Sigma_{hh}}$ can be obtained from actual channel measurements, or can be numerically estimated by measuring the correlation over many realizations of a given channel model). As noted in Section~\ref{sec:channelmodel}, the channel models for OFDM-based UWB communication satisfy this assumption.

\SubSubSection{Average PEP}

Noting that only the $\eta_{i,\ell}$ non-zero terms of $(\ve{x}-\ve{z}_{i,\ell})$ in (\ref{eq:PEP}) contribute to the PEP (and suppressing the dependence of $\eta$ on $i$ and $\ell$ for notational clarity), we let $\ve{x}'$, $\ve{z}'_{i,\ell}$, $\ve{H}'=\diag(\ve{h}')$, $\ve{J}'$, and $\ve{n}'$ represent the transmitted symbols, received symbols, channel gains, interferences, and AWGN noises corresponding to the $\eta$ non-zero entries of $(\ve{x}-\ve{z}_{i,\ell})$, respectively, and form $\ve{\Sigma_{h'h'}}$ by extracting the the elements from $\ve{\Sigma_{hh}}$ which correspond to $\ve{h}'$. Letting $\ve{D} = \diag(\ve{x}'-\ve{z}'_{i,\ell})$ be the diagonal matrix of non-zero entries and $\ve{g} = \ve{H}'(\ve{x}'-\ve{z}'_{i,\ell}) =  \ve{D}\ve{h}'$, we have 
\begin{eqnarray}
\Ex(\ve{g}) &=& \ve{0}_{\eta{\times}1} \;, \\
\Ex(\ve{g}\ve{g}^H) &=& \ve{R_{gg}} = \ve{D}\ve{\Sigma_{h'h'}}\ve{D}^H \;,
\end{eqnarray}
i.e., the distribution of $\ve{g}$ is zero-mean complex Gaussian with covariance matrix $\ve{R_{gg}}$. 

We would like to obtain the average $\overline{\PEP}_{i,\ell}$ for the $\ell$th error vector, starting in the $i$th position. Rewriting (\ref{eq:PEP}) including only the contributing terms, we obtain 
\begin{eqnarray}
\overline{\PEP}_{i,\ell} &=& \mathrm{Pr}\left\{||\ve{r}'-\ve{H}'\ve{z}'_{i,\ell}||^2 - ||\ve{r}' - \ve{H}'\ve{x}'||^2 < 0 \right\} \;, \\
 &=& \mathrm{Pr}\left\{ \ve{g}\ve{g}^H - \ve{g}(\ve{J}'+\ve{n}')^H - (\ve{J}'+\ve{n}')\ve{g}^H < 0 \right\} \;, \nonumber \\
 &=& \mathrm{Pr}\left\{ \Delta_{i,\ell}(\ve{D}) < 0 \right\} \;,
\end{eqnarray}
where $\Delta_{i,\ell}(\ve{D}) = \ve{y}^H\ve{A}\ve{y}$ and
\begin{displaymath}
\ve{y} = \left[\!\!\!\begin{array}{c}\ve{g} \\ %
                                  \ve{J}' + \ve{n}' \end{array}\!\!\!\right] \;, %
\quad\quad\quad\quad%
\ve{A} = \left[\!\!\!\begin{array}{cc} \ve{I}_\eta  & -\ve{I}_\eta \\ %
                                  -\ve{I}_\eta & \ve{0}_\eta \end{array}\!\!\!\right] \;. %
\end{displaymath}

We adopt the Laplace transform approach \cite{BCTV98} to determine $\mathrm{Pr}\left\{ \Delta_{i,\ell}(\ve{D}) < 0 \right\}$, and consider two typical narrowband channel situations:

\textsl{Case 1 --- $\alpha_k$ constant (non-faded interferers):} In this case, we note $\ve{y}$ has mean $\ve{\mu_{yy}}$ and covariance matrix $\ve{R_{yy}}$, which are given by
\begin{displaymath}
\ve{\mu_{yy}} = \Ex(\ve{y}) =  \left[\!\!\!\begin{array}{c}\ve{0}_{\eta{\times}1} \\ %
                                  \ve{J}' \end{array}\!\!\!\right] \;, %
\quad\quad\quad\quad %
\ve{R_{yy}} = \left[\!\!\!\begin{array}{cc} \ve{R_{gg}} & \ve{0}_\eta \\ %
                                  \ve{0}_\eta            & \No\ve{I}_\eta \end{array}\!\!\!\right] \;.
\end{displaymath}

The Laplace transform of $\Delta_{i,\ell}(\ve{D})$ is given by~\cite{SBS66}
\begin{equation}
\label{eq:phiAWGN}
\Phi_{i,\ell}(s) = %
\frac{\exp[-s\ve{\mu}^{H}_{\ve{yy}}(\ve{A}^{-1} + s\ve{R_{yy}})^{-1}\ve{\mu_{yy}}]} %
               {\det(\ve{I}_{2\eta} + s\ve{R_{yy}A})} \;.
\end{equation}

\textsl{Case 2 --- $\alpha_k$ independent Rayleigh faded interferers:} In this case, $\Ex(\ve{y})=\ve{0}_{2\eta{\times}1}$, and we have
\begin{equation}
\ve{R_{yy}} = \left[\!\!\!\begin{array}{cc} \ve{R_{gg}} & \ve{0}_\eta \\ %
                                  \ve{0}_\eta            & \ve{R_{J'J'}} + \No\ve{I}_\eta \end{array}\!\!\!\right] \;,
\end{equation}
where $\ve{R_{J'J'}} = \Ex\;(\ve{J}'\ve{J}'^H)$, and the Laplace transform of $\Delta_{i,\ell}(\ve{D})$ is given by
\begin{equation}
\label{eq:phiRayl}
\Phi_{i,\ell}(s) = %
\frac{1}{\det(\ve{I}_{2\eta} + s\ve{R_{yy}A})} \;.%
\end{equation}

In either case, the average PEP for the $\ell$th error vector starting in the $i$th position is given by~\cite{BCTV98}
\begin{equation}
\label{eq:PEPavg}
\overline{\PEP}_{i,\ell} = \mathrm{Pr}\left\{ \Delta_{i,\ell}(\ve{D}) < 0 \right\} = %
\frac{1}{2{\pi}j}\int\limits_{c-j\infty}^{c+j\infty} %
\Phi_{i,\ell}(s) \frac{\mathrm{d}s}{s} \;,
\end{equation}
where $c$ is in the convergence region of $\Phi_{i,\ell}(s)$. We note that (\ref{eq:PEPavg}) may be solved efficiently either analytically or via numerical integration using a Gauss-Chebyshev quadrature rule~\cite{BCTV98}.

\SubSubSection{Average PEP without Interference}
\label{sec:avgpep_nointfr}

An alternative form for the average PEP can be obtained for the special case of $\ve{J}=\ve{0}_{N{\times}1}$. From (\ref{eq:PEPnoI}), and following \cite[Eq. (7)]{V01}, we can write the average PEP for the $\ell$th error vector starting in the $i$th position as 
\begin{equation}
\label{eq:ProbInt}
\overline{\PEP}_{i,\ell} = \frac{1}{\pi}\int\limits_{0}^{\pi/2}\left[ \det\left( \frac{E_s\ve{R_{gg}}}{\No\sin^2\theta} + \ve{I}_\eta\right) \right]^{-1}d\theta\;.
\end{equation}
It can be shown that the Laplace transform approach with $\ve{J}=\ve{0}_{N{\times}1}$ leads to an equivalent expression.

\SubSubSection{Average BER}

Given the average PEP according to either (\ref{eq:PEPavg}) or (\ref{eq:ProbInt}), the corresponding bit error rate for the $\ell$th error vector, starting in the $i$th position, is given by%
\begin{equation}
\label{eq:Pil_avg}
\bar{P}_{i,\ell} = a_\ell \cdot \overline{\PEP}_{i,\ell} \;.
\end{equation}

Summing over all $L$ error vectors, the BER for the $i$th starting position can be written as
\begin{equation}
\bar{P}_i = \sum_{\ell=1}^{L}\bar{P}_{i,\ell} \;.
\end{equation}
Finally, since all starting positions are equally likely to be used, the average BER $\bar{P}$ can be written as 
\begin{equation}
\label{eq:BERavg}
\bar{P} = \frac{1}{L_c}\sum_{i=1}^{L_c}\bar{P}_i = \frac{1}{L_c}\sum_{i=1}^{L_c} \sum_{\ell=1}^{L} \bar{P}_{i,\ell} \;.
\end{equation}
Table~\ref{table:pseudo} contains pseudocode to calculate $\bar{P}$ according to (\ref{eq:BERavg}). Note that, since $\bar{P}_{i,\ell}$ in (\ref{eq:BERavg}) is already averaged over $\ve{H}$, we cannot upper-bound it by $1/2$ as we did in (\ref{eq:BERgivenH}) for Method~I. This implies that the result for Method~II may be somewhat looser than that for Method~I (see also Section~\ref{sec:no_interference}).

\Section{Numerical Results}
\label{sec:results}

In this section, we present numerical results for Methods I and II introduced above. We focus on the particular case of MB-OFDM operating in the CM1 UWB channel. As mentioned in Section~\ref{sec:channelmodel}, for Method II we include the effect of ``outer'' lognormal shadowing by numerically integrating the results of (\ref{eq:BERavg}) over the appropriate lognormal distribution~\cite{MolischEtAl03}.

\SubSection{No Interference}
\label{sec:no_interference}

\begin{figure}[t]
\centerline{\resizebox{0.9\columnwidth}{!}{\input{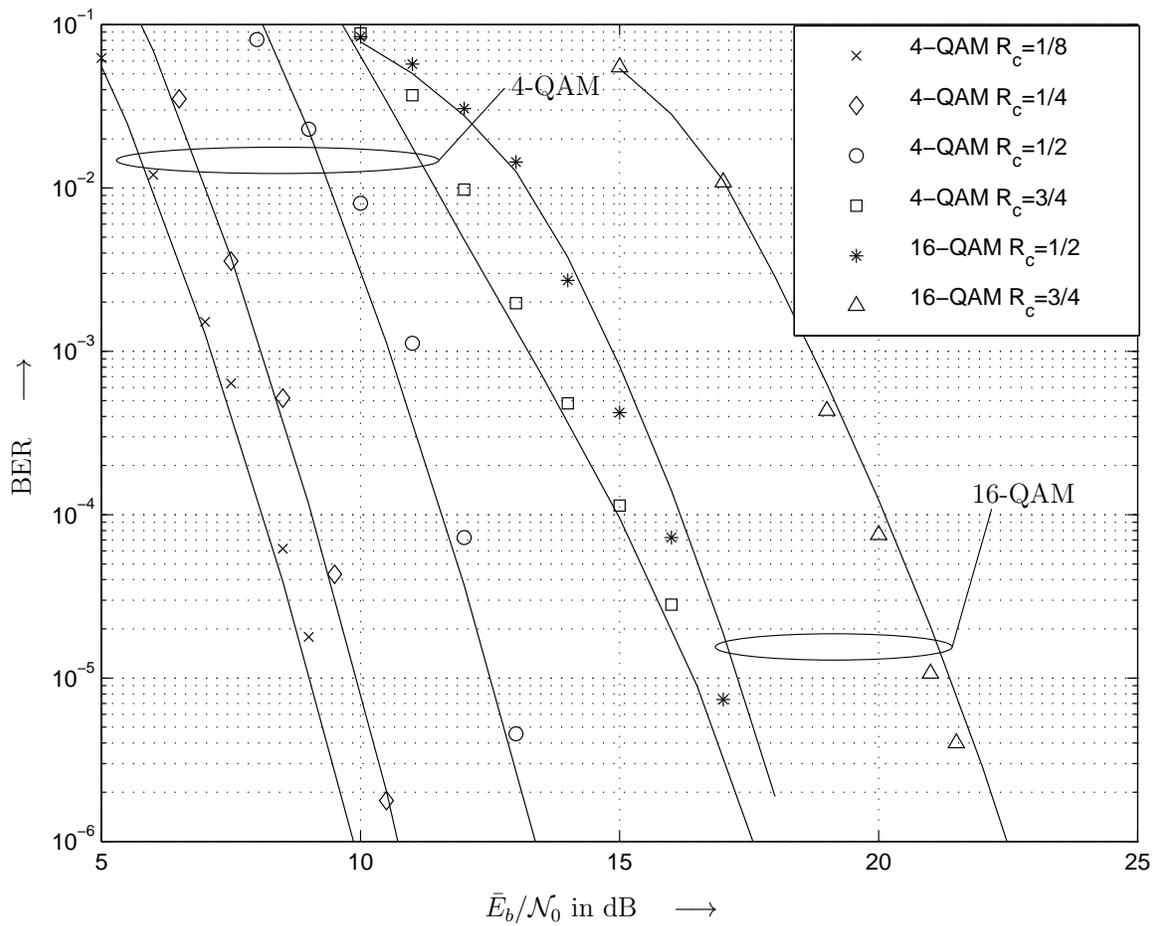}}}
\caption{\label{fig:realiz}10\% outage BER vs.\ $\SNR$ from Method I (lines) and simulation results (markers) for different code rates and modulation schemes. UWB CM1 channel. Code rates $1/4$ and $1/8$ include repetition. No interference ($\ve{J}=\ve{0}_{N{\times}1}$).}
\end{figure}

In Figure~\ref{fig:realiz}, we present the 10\% outage BER as a function of $\SNR$ obtained using Method I (lines), as well as simulation results (markers) for different code rates and modulation schemes using a set of 100 UWB CM1 channel realizations with lognormal shadowing, where $\bar{E}_b$ denotes the mean received energy per information bit over the ensemble of channels. The 10\% outage BER is a common performance measure in UWB systems, cf. e.g.~\cite{MBOFDM}. We can see that Method I is able to accurately predict the outage BER for 4-QAM and 16-QAM modulation schemes and a variety of different code rates, with a maximum error of less than 0.5~dB. It is also important to note that obtaining the Method I result requires significantly less computation than is required to obtain the simulation results for all 100 UWB channel realizations. For example, it took about 15 minutes to obtain one of the analytical curves of Figure~\ref{fig:realiz}, while it took approximately 48 hours to obtain the corresponding simulation results on the same computer.

\begin{figure}[t]
\centerline{\resizebox{0.9\columnwidth}{!}{\input{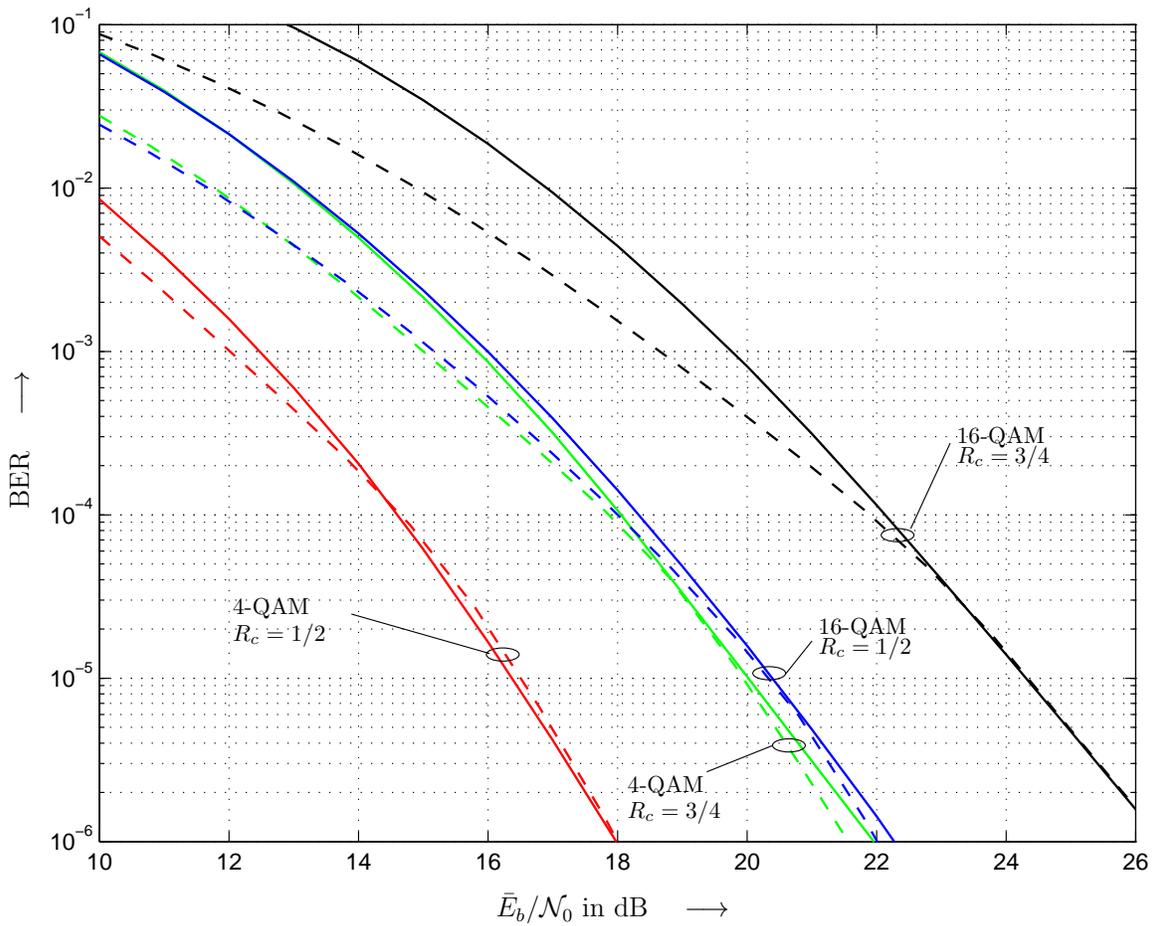}}}
\caption{\label{fig:compareMethods} Average BER versus $\SNR$ for 4-QAM and 16-QAM with code rates $R_c=1/2$ and $3/4$. Solid lines: Direct average from Method II. Dashed lines: Method I with an average over 10,000 channel realizations. UWB CM1 channel. No interference ($\ve{J}=\ve{0}_{N{\times}1}$).}
\end{figure}

Figure~\ref{fig:compareMethods} illustrates the average BER as a function of $\SNR$ for 4-QAM and 16-QAM with code rates $R_c=1/2$ and $3/4$ using two approaches: Method I with an average over 10,000 channel realizations (dashed lines), and the direct average from Method II (solid lines). As expected, the two methods are in close agreement at low BER. The deviation between the two results at higher BER is due to (a) the loosening effect of the averaging of Method II over the lognormal distribution, and (b) the fact that Method I is somewhat tighter due to the upper-bounding by $1/2$ in (\ref{eq:BERgivenH}).

\begin{figure}[t]
\centerline{\resizebox{0.9\columnwidth}{!}{\input{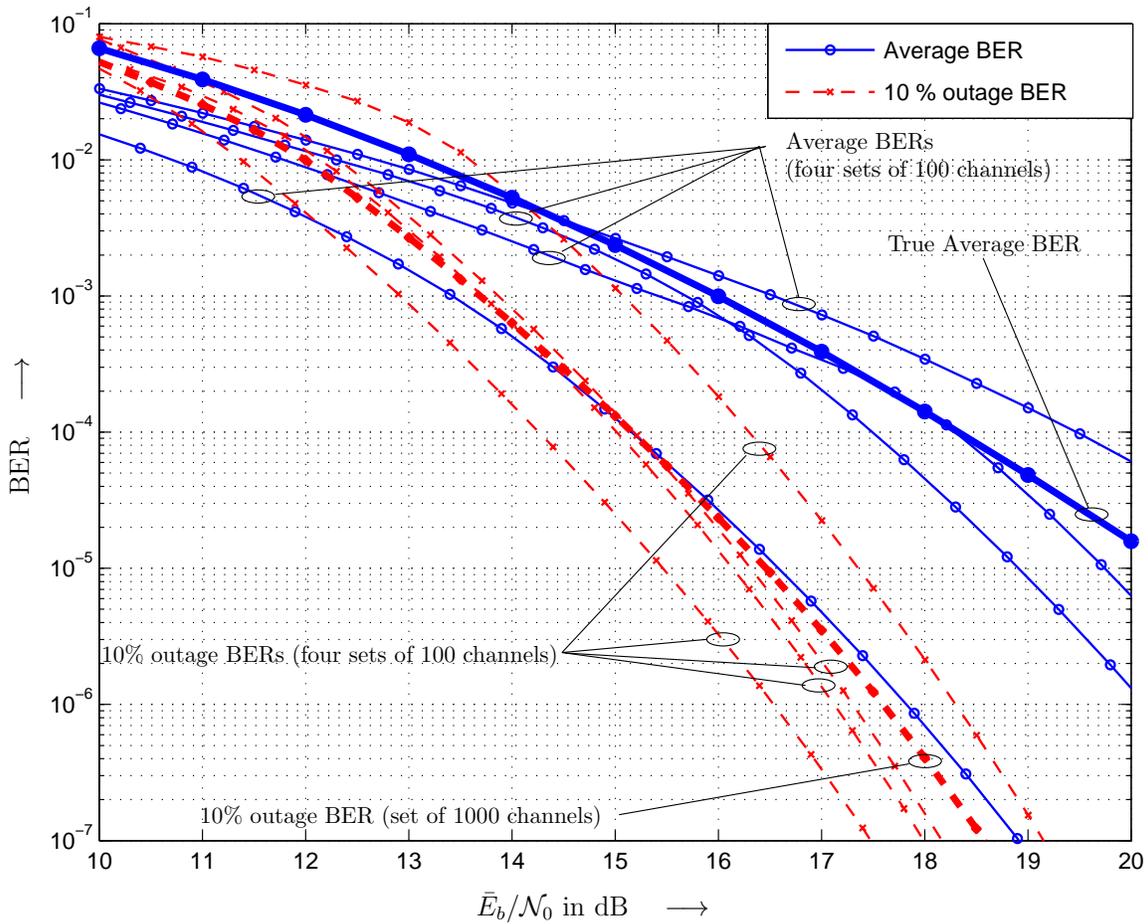}}}
\caption{\label{fig:diff_avgandout}Average BER (solid lines) and 10\% outage BER (dashed lines) versus $\SNR$ for four different sets of 100 channels using Method I. For comparison: average BER from Method II (bold solid line), and 10\% outage BER for a set of 1000 channels (bold dashed line). UWB CM1 channel, $R_c=1/2$, 16-QAM. No interference ($\ve{J}=\ve{0}_{N{\times}1}$).}
\end{figure}

\textsl{A Caution to System Designers:} We should note that 100 channel realizations (standard for MB-OFDM performance analysis \cite{BatraEtAl04}) may not be sufficient to accurately capture the true system performance. Figure~\ref{fig:diff_avgandout} (solid lines) shows the average BER with respect to $\SNR$ for four different sets of 100 UWB CM1 channel realizations, obtained via Method I. For comparison, the average performance obtained via Method II is also shown (bold solid line). We can see that the average system performance obtained using sets of only 100 channel realizations depends greatly on the specific realizations which are chosen. Similarly, Figure~\ref{fig:diff_avgandout} illustrates the 10\% outage BER with respect to $\SNR$ for four different sets of 100 UWB CM1 channel realizations, obtained via Method I (dashed lines). For comparison the 10\% outage BER obtained using a set of 1000 realizations is also shown (bold dashed line). We see that the outage BER curves, while less variable than the average BER curves, are still quite dependent on the selected channel realization set.

Based on the results above, it seems that performance evaluation for systems operating in quasi-static channels using only small numbers of channel realizations may be prone to inaccurate results. This is one of the main strengths of the two methods presented in Section~\ref{sec:analysis}: the performance can easily be evaluated over any number of channel realizations (Method I), or the average performance can be directly obtained (Method II), without resorting to lengthy simulations.

\SubSection{Non-Faded Tone Interference}
\label{sec:results_awgnintf}

In this section, we present results for non-faded tone interference, specifically focusing on the MB-OFDM system operating at 320~Mbps ($R_c=1/2$ after puncturing) with 4-QAM modulation over the CM1 channel. We concentrate on the case of $N_i=1$ interferer, in order to examine the effect of the interferer frequency $f_1$ and the signal-to-interference ratio\footnote{\sf Note that the $\SIR$ according to this definition is an average over all the subcarriers, so the $\SIR$ for a specific subcarrier may be much higher/lower than the average. For example, in the 384-subcarrier MB-OFDM system with one interferer directly on a subcarrier, the $\SIR$ of the affected subcarrier will be $\approx 26$ dB lower than the average $\SIR$ (since the interference on all other subcarriers is zero).} 
\begin{equation}
\SIR = \frac{\Ex\,(||\ve{H}\ve{x}||^2)}{\Ex\,(||\ve{J}||^2)} \;.
\end{equation}
Without loss of generality we place $f_1$ between the $52^{\mathrm{nd}}$ and $53^{\mathrm{rd}}$ MB-OFDM subcarriers.

\begin{figure}[t]
\centering \resizebox{0.9\columnwidth}{!}{\input{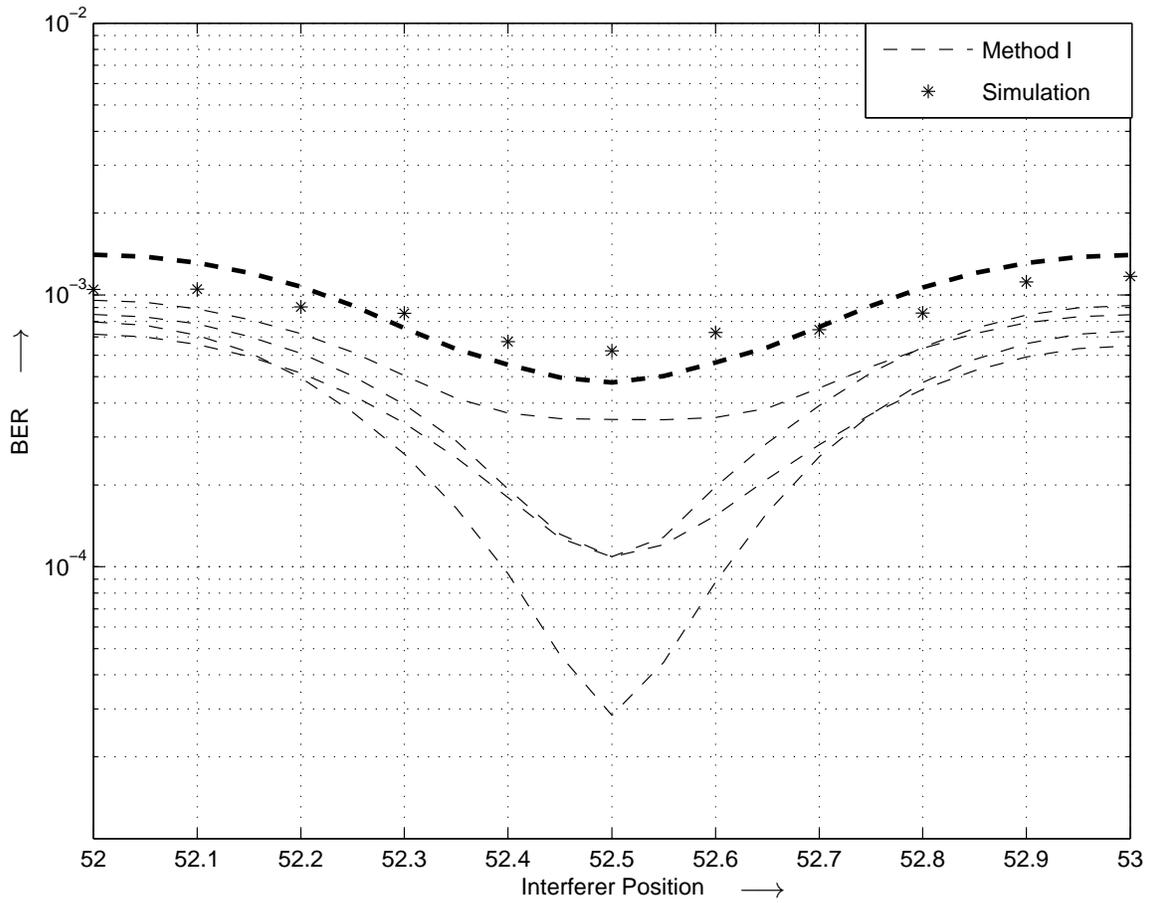}}
\caption{\label{fig:BERvsPOS_awgnIntf_both}Average BER versus interferer position for $\SNR = 17$ dB and $\SIR = 19$ dB with one non-faded interferer. Results shown: Method I with different sets of 100 channel realizations averaged over 32 phases $\phi_1\in[0,2\pi)$ (dashed lines); simulation results for 100 channel realizations corresponding to bold dashed line (markers). UWB CM1 channel, $R_c=1/2$, 4-QAM.}
\end{figure}

In Figure~\ref{fig:BERvsPOS_awgnIntf_both}, we consider $\SNR = 17$ dB, $\SIR = 19$ dB, and focus on the effect of varying $f_1$. We show the average BER for five different sets of 100 channel realizations (dashed lines), obtained via Method I. The markers ($\ast$) indicate simulation results which correspond to, and are in good agreement with, the Method~I results for the set of 100 channels indicated by a bold dashed line. Figure~\ref{fig:BERvsPOS_awgnIntf_both} indicates that the best-case performance is obtained when $f_1$ lies exactly between two OFDM subcarriers (interferer position 52.5), and the performance degrades as $f_1$ approaches a subcarrier frequency.  We also note that, as seen in Figure~\ref{fig:diff_avgandout} for the no-interference case, the performance obtained using Method I can be quite variable when considering small sets of channel realizations.

Figure~\ref{fig:BERvsSNR_awgnIntf} shows the BER versus $\SNR$ for one non-faded interferer at positions 52.0 (solid lines) and 52.5 (dashed lines) with $\SIR=\{28,23,21,19,17,15\}$ dB, obtained using Method~II. For comparison, the no-interference ($\SIR=\infty$) performance from Method II (bold solid line) is also shown. This figure clearly illustrates the performance degradation associated with decreasing $\SIR$. As seen in Figure~\ref{fig:BERvsPOS_awgnIntf_both}, the best-case performance is obtained when $f_1$ lies exactly between two OFDM subcarriers, while the performance degrades as $f_1$ approaches a subcarrier frequency.

\SubSection{Rayleigh-faded Tone Interference}
\label{sec:results_raylintf}

We now consider the effect of $N_i=1$ Rayleigh-faded interferer, in order to compare the relative effects of interference with those of the non-faded interferer in Section~\ref{sec:results_awgnintf}. Figure~\ref{fig:BERvsSNR_raylIntf} shows the BER versus $\SNR$ obtained using Method II for the same interferer positions and $\SIR$ values as in Figure~\ref{fig:BERvsSNR_awgnIntf}. By comparing Figures \ref{fig:BERvsSNR_awgnIntf} and \ref{fig:BERvsSNR_raylIntf}, we can clearly see that Rayleigh-faded tone interferers have a larger effect on the BER performance than non-faded tone interferers. For example, at $\SNR = 17$~dB, $\SIR = 21$~dB and interferer position 52.5, the BER with one non-faded interferer is approximately $10^{-5}$, while the BER with one Rayleigh interferer is approximately $2.3{\times}10^{-4}$. Even for relatively high $\SIR=28$~dB, one Rayleigh tone interferer at position 52.0 causes a much larger effect than the non-faded tone interferer at the same $\SIR$.

\SubSection{Interference Mitigation by Erasure Marking and Decoding}
\label{sec:results_erasures}

In OFDM systems where interference impacts a small number of subcarriers, one simple and practical method of interference mitigation is to erase the information bits carried on the most-affected subcarriers (proposed in e.g. \cite{DP02}, as well as more advanced joint marking and decoding in \cite{LMS03}). In order to study the potential performance of such an erasure technique, we consider the use of a genie which erases the subcarriers with largest interference powers. In the framework of analysis of Section~\ref{sec:analysis}, subcarrier erasures can be considered as additional puncturing and easily incorporated into both analysis methods.

\begin{figure}[t]
\centering \resizebox{0.9\columnwidth}{!}{\input{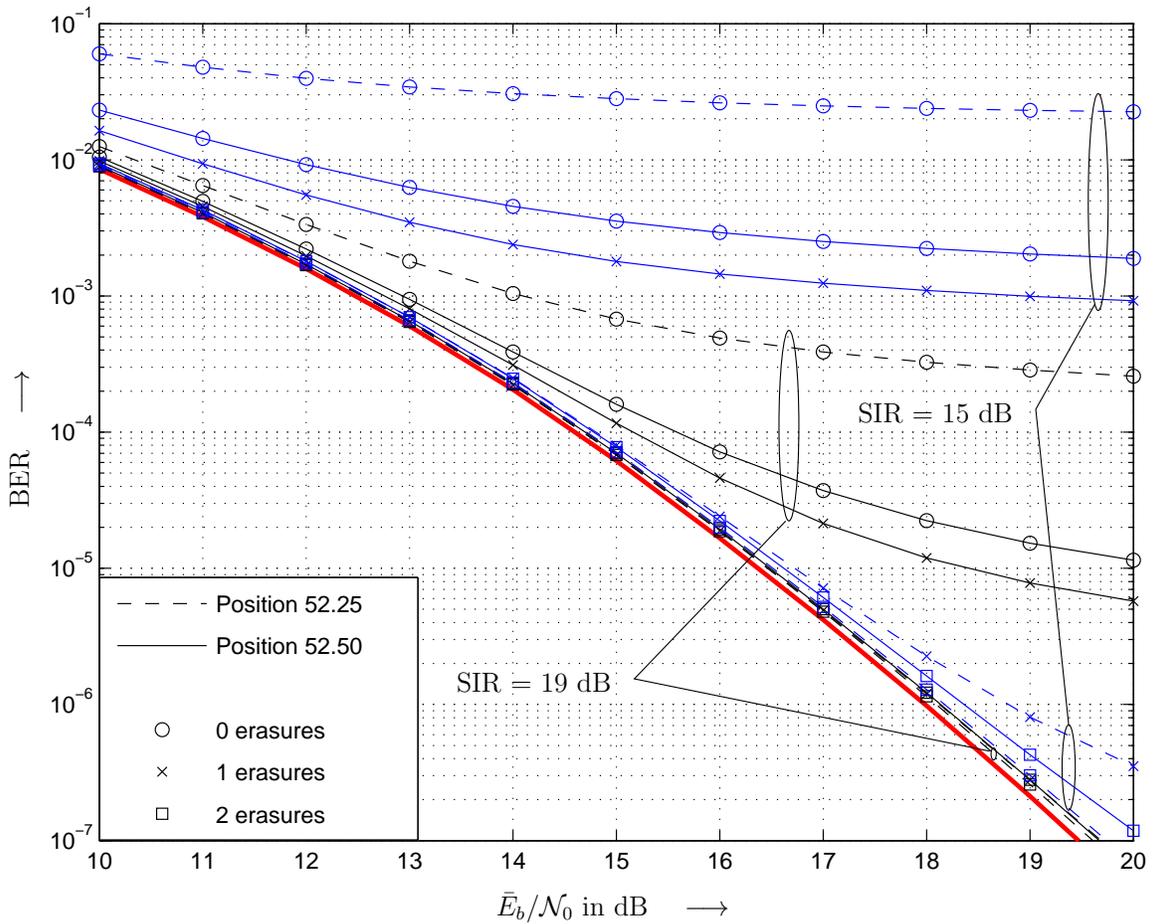}}
\caption{\label{fig:erasures_awgn}Average BER versus $\SNR$ for $\{0,1,2\}$ subcarrier erasures. One non-faded interferer, positions 52.25 (dashed lines) and 52.5 (dashed lines) and $\SIR$ = $\{15,19\}$ dB, obtained using Method II. For comparison: $\SIR=\infty$ ($\ve{J}=\ve{0}_{N{\times}1}$) from Method II (bold solid line). UWB CM1 channel, $R_c=1/2$, 4-QAM.}
\end{figure}

Figure \ref{fig:erasures_awgn} illustrates the average BER versus $\SNR$ for 0, 1, and 2 subcarrier erasures, obtained using Method~II. One non-faded interferer is placed at positions 52.25 (dashed lines) and 52.5 (solid lines), with $\SIR$ of 15 and 19 dB. As can be seen from this figure, the use of a small number of subcarrier erasures rapidly decreases the effect of the tone interference and allows the interference-impaired system performance to approach the no-interference performance (bold solid line). Focusing on the case of position 52.5 (solid lines), we can see that using only one erasure has a small effect on the resultant BER. This is due to the windowing effect of the DFT at the OFDM receiver (see Section~\ref{sec:interferencemodel}), which results in interfering signal power being symmetrically distributed amongst a number of subcarriers. However, once the two largest equal-interference-power subcarriers are erased, performance improves dramatically. On the other hand, when the tone interferer is at position 52.25, a large portion of the interference power is in one subcarrier, so even one erasure can provide a substantial performance improvement. We should also note that if the interferer happens to be exactly at the subcarrier frequency, one subcarrier erasure will suffice to totally remove the effect of the interference.

\begin{figure}[t]
\centering \resizebox{0.9\columnwidth}{!}{\input{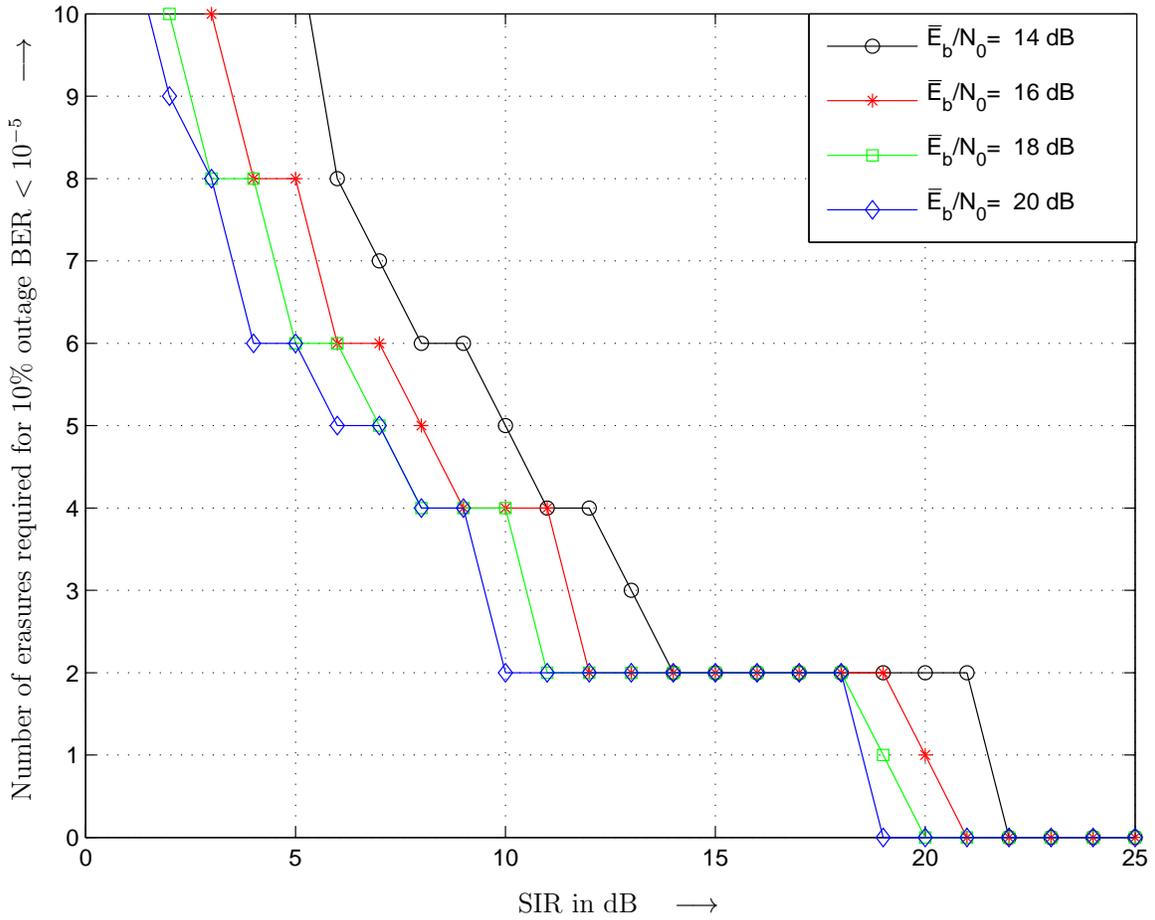}}
\caption{\label{fig:erasures_awgn_outage}Required number of subcarrier erasures to maintain 10\% outage BER $< 10^{-5}$ for various $\SIR$ and $\SNR$. One non-faded interferer, position 52.5, average over 32 phases $\phi_1\in[0,2\pi)$, obtained using Method I with 1000 UWB CM1 channel realizations, $R_c=1/2$, 4-QAM.}
\end{figure}

We conclude by returning once again to the consideration of outage BER obtained via Method~I. In Figure \ref{fig:erasures_awgn_outage}, we consider one non-faded interferer at position 52.5, and show the number of subcarrier erasures required to maintain the 10\% outage BER $< 10^{-5}$ for varying $\SIR$ and different values of $\SNR$. As expected, decreasing the $\SIR$ results in a higher required number of erasures to maintain the target BER. Unfortunately, a large number of erasures compromise the code's error correcting capability. As can be seen from Figure~\ref{fig:erasures_awgn_outage}, eventually too many erasures weaken the code sufficiently that, even with the effects of interference mostly removed, the code is not able to maintain the required target BER. Figure~\ref{fig:erasures_awgn_outage} also shows that providing an increased SNR margin allows the MB-OFDM system to compensate for a larger amount of interference.

\Section{Conclusions}
\label{sec:conclusions}

In this paper, we have presented two methods for evaluating the performance of convolutionally-coded multicarrier systems employing $M$-QAM, operating over frequency-selective, quasi-static, non-ideally interleaved fading channels and impaired by sum-of-tones interference.
The realization-based method (``Method I'') presented in Section~\ref{sec:rba} estimates the system performance for each realization of the channel with an arbitrary fading distribution, and is suitable for evaluating the outage performance of systems.  The method presented in Section~\ref{sec:avg} (``Method II''), based on knowledge of the correlation matrix of the Rayleigh-distributed frequency-domain channel gains, allows for direct calculation of the average system performance over the ensemble of quasi-static fading channel realizations. 

These two novel methods allow for analytical evaluation of the performance of a class of systems of considerable practical interest, whose performance evaluation was previously only possible via intensive numerical simulations. The results in Section~\ref{sec:results} demonstrate that the proposed methods of analysis provide an accurate measure of the system performance and allow for much greater flexibility than simulation-based approaches.

We have also shown that the MB-OFDM system (and OFDM systems in general) may be significantly impacted by the effect of tone interference, but that this performance degradation can be mitigated to a large extent by the use of erasure marking and decoding at the receiver, provided that the receiver can obtain knowledge of which subcarriers are impaired by the interferers.

\renewcommand{\baselinestretch}{1.0}\normalsize\makeatletter
\def\@listi{\leftmargin\leftmargini
            \parsep 5\p@  \@plus2.5\p@ \@minus\p@
            \topsep 10\p@ \@plus4\p@   \@minus6\p@
            \itemsep-5\p@  \@plus1\p@ \@minus0\p@}
\makeatother
\bibliography{IEEEabrv,local}
\bibliographystyle{IEEEtran}

\renewcommand{\baselinestretch}{1.05}\normalsize

\newpage

\textbf{Tables and Figures:}
\vspace{5mm}

\begin{table}[h!]
\caption{\label{table:pseudo}Pseudocode for the two analysis methods.}
\vspace{2mm}
\centerline{\framebox[4.65in][c]{\begin{tabular}{|ll|}
\hline
{\sf Method I} & {\sf Final BER is $P$ (for given $\ve{H}$, $\ve{J}$).} \\
\hline
\tt 1  & \tt $P$ := 0\\
\tt 2  & \tt for $i$ := 1 to $L_c$ do \\
\tt 3  & \tt ~~$P_i$ := 0 \\
\tt 4  & \tt ~~for $\ell$ := 1 to $L$ do \\
\tt 5  & \tt ~~~~form $\ve{q}_{i,\ell}$ as per (\ref{eq:q})\\
\tt 6  & \tt ~~~~form $\ve{v}_{i,\ell}$ as per (\ref{eq:v})\\
\tt 7  & \tt ~~~~form $\ve{v}^\pi_{i,\ell}$ and $\ve{z}_{i,\ell}$ from $\ve{v}_{i,\ell}$\\
\tt 8  & \tt ~~~~calculate $\PEP_{i,\ell}$ as per (\ref{eq:PEPij})\\
\tt 9  & \tt ~~~~calculate $P_{i,\ell}$ as per (\ref{eq:BERij})\\
\tt 10  & \tt ~~~~$P_i$ := $P_i$ + $P_{i,\ell}$\\
\tt 11  & \tt ~~endfor \\
\tt 12 & \tt ~~$P$ := $P$ + min($\frac{1}{2}$,$P_i$)\\
\tt 13 & \tt endfor \\
\tt 14 & \tt $P$ := $P$ / $L_c$ \\
\hline
\hline
{\sf Method II} & {\sf Final BER is $\bar{P}$.} \\
\hline
\tt 1  & \tt $\bar{P}$ := 0\\
\tt 2  & \tt for $i$ := 1 to $L_c$ do \\
\tt 3  & \tt ~~for $\ell$ := 1 to $L$ do \\
\tt 4  & \tt ~~~~form $\ve{q}_{i,\ell}$ as per (\ref{eq:q}) \\
\tt 5  & \tt ~~~~form $\ve{v}_{i,\ell}$ as per (\ref{eq:v})\\
\tt 6  & \tt ~~~~form $\ve{v}^\pi_{i,\ell}$ and $\ve{z}_{i,\ell}$ from $\ve{v}_{i,\ell}$\\
\tt 7  & \tt ~~~~form $\ve{x'}_{i,\ell}$, $\ve{z'}_{i,\ell}$, $\ve{h'}$, $\ve{J'}$ and $\ve{\Sigma_{h'h'}}$ \\
\tt 8  & \tt ~~~~compute $\ve{D} := \diag(\ve{x'}-\ve{z'}_{i,\ell})$ \\
\tt 9  & \tt ~~~~compute $\ve{g}=\ve{D}\ve{h'}$ and $\ve{R_{gg}} := \ve{D}\ve{\Sigma_{h'h'}}\ve{D}^H$ \\
\tt 10 & \tt ~~~~form $\ve{\mu}_\ve{yy}$ and/or $\ve{R_{J'J'}}$ (as required) \\
\tt 11 & \tt ~~~~form $\ve{R}_\ve{yy}$ and $\ve{A}$ \\
\tt 12 & \tt ~~~~form $\Phi_{i,\ell}(s)$ as per either (\ref{eq:phiAWGN}) or (\ref{eq:phiRayl}) \\
\tt 13 & \tt ~~~~calculate $\overline{\PEP}_{i,\ell}$ as per (\ref{eq:PEPavg}) \\
\tt 14 & \tt ~~~~calculate $\bar{P}_{i,\ell}$ as per (\ref{eq:Pil_avg}) \\
\tt 15 & \tt ~~~~$\bar{P}$ := $\bar{P}$ + $\bar{P}_{i,\ell}$\\
\tt 16 & \tt ~~endfor \\
\tt 17 & \tt endfor \\
\tt 18 & \tt $\bar{P}$ := $\bar{P}$ / $L_c$ \\
\hline
\end{tabular}}}
\end{table}

\renewcommand{\baselinestretch}{1.15}\normalsize
\clearpage

\clearpage

\clearpage

\clearpage

\clearpage

\clearpage
\begin{figure}[t]
\centering \resizebox{0.9\columnwidth}{!}{\input{figures/awgnintfr_berVSsnr.pstex_t}}
\caption{\label{fig:BERvsSNR_awgnIntf}Average BER versus $\SNR$ for various $\SIR$ with one non-faded interferer, obtained using Method II. Interferer positions 52.0 (solid lines) and 52.5 (dashed lines). For comparison: $\SIR=\infty$ ($\ve{J}=\ve{0}_{N{\times}1}$) from Method II (bold solid line). UWB CM1 channel, $R_c=1/2$, 4-QAM.}
\end{figure}

\clearpage
\begin{figure}[t]
\centering \resizebox{0.9\columnwidth}{!}{\input{figures/raylintfr_berVSsnr.pstex_t}}
\caption{\label{fig:BERvsSNR_raylIntf}Average BER versus $\SNR$ for various $\SIR$ with one Rayleigh-faded interferer, obtained using Method II. Interferer positions 52.0 (solid lines) and 52.5 (dashed lines). For comparison: $\SIR=\infty$ ($\ve{J}=\ve{0}_{N{\times}1}$) from Method II (bold solid line). UWB CM1 channel, $R_c=1/2$, 4-QAM.}
\end{figure}

\clearpage

\clearpage

\end{document}